%
%

\input epsf

\def\mpfig#1{PS#1.ps}

\font\csc=cmcsc10 
\def\caption Figure #1. {{\csc Figure #1}.\enspace}

\magnification\magstephalf
\baselineskip14pt
\parskip3pt

\def\mathcal#1{\cal{#1}}

\def\pfbox
  {\hbox{\hskip 10pt\lower2pt\vbox{\hrule
  \hbox to 7pt{\vrule height 5pt\hfill\vrule}
  \hrule}}\hskip3pt}

\def\bib[#1] {\par\noindent\hangindent 40pt\hbox to20pt{[#1]\hfil}}

\def\dilog{\mathop{\rm Li_2}}
\def\PS{\mathop{\rm PS_3}}
\def\frac#1#2{{\textstyle{#1\over#2}}}
\def\eps{\varepsilon}
\def\sxi{\sqrt{\xi}}
\def\lxi{\ln\xi}

\def\Byckling{1}
\def\Sterman{2}
\def\PhD{3}
\def\NPB{4} 
\def\Lewin{5}

\centerline{\bf Closed-form analytic solutions for dilogarithmic double integrals}
\bigskip
\centerline{\sl Michael M.~Tung\footnote{$^1$}{\rm email: mtung@mat.upv.es}
and Lucas J\'odar\footnote{$^2$}{\rm email: ljodar@mat.upv.es}}
\centerline{\sl Instituto de Matem\'atica Multidisciplinar}
\centerline{\sl Universidad Polit\'ecnica de Valencia}
\centerline{\sl P.O.~Box 22.012, Valencia, Spain}
\bigskip\bigskip
{\narrower\smallskip\noindent
{\bf Abstract.} 
This note presents techniques to analytically solve double integrals of the
dilogarithmic type which are of great importance in the perturbative treatment
of quantum-field theory. In our approach divergent integrals can be calculated
similar to their convergent counterparts after identifying and isolating their
singular parts.
\smallskip}
{\narrower\smallskip\noindent
{\bf Keywords:} 
integration method, dilogarithmic integral
\smallskip}

\bigskip\noindent
{\bf 1. Introduction and physical motivation.}\enspace
In quantum field theory the differential volume element for the three-particle
phase space in four-dimensional Minkowski space has the form (neglecting
overall normalization and statistical factors) of a product containing the
relativistic four-momenta $p_i$ of the three particles:
$$d\PS\sim\prod\limits_{i=1}^3 d^4\!p_i.\eqno(1)$$
Usually one includes energy-momentum conservation
($q=\sum_{i=1}^3p_i$) and the on-shell requirement for the particles
($p_i^2=m_i^2$) as Dirac delta functions with positive-energy solutions
so that
$$d\PS\sim\prod\limits_{i=1}^3 d^4\!p_i\;\delta_+\!(p_i^2-m_i^2)
       \;\delta^4\!\left(q-\sum\limits_{i=1}^3p_i\right),\eqno(2)$$
where $q$ is the total momentum of the reaction and $m_i$ are the masses
of the final particles under consideration.
This expression is known as the Lorentz invariant three-body phase space
and its integration over the kinematic $S$-matrix yields numerical
predictions for the measurable cross sections in physical experiments
(see {\it e.g.\/} [\Byckling] and [\Sterman]). 

The general phase space Eq.~(2) has $5$ degrees of freedom, resulting from
$7$ restrictions ($3$ by the on-shell requirement and $4$ by energy-momentum
conservation) on the overall $12$ components of the four-momenta.
Integrating out all dependence of the undetected spatial components of
momentum $p_3$, leaves only two degrees of freedom to parameterize the
phase space. Therefore, a convenient choice to describe the phase space are
the two Lorentz invariants~[\PhD]:
$$y=1-{2p_1\!\cdot q\over q^2}\quad\hbox{and}\quad
z=1-{2p_2\!\cdot q\over q^2}.\eqno(3)$$

\def\\#1#2{\vtop{\halign{\hfil##\hfil\cr
 \def\epsfsize##1##2{.8##1}\epsfbox{\mpfig#1}\cr
 \noalign{\smallskip}
 (#2)\cr
}}}

\line{\hfil\\1a\hfil\\2b\hfil\\3c\hfil}
\medskip
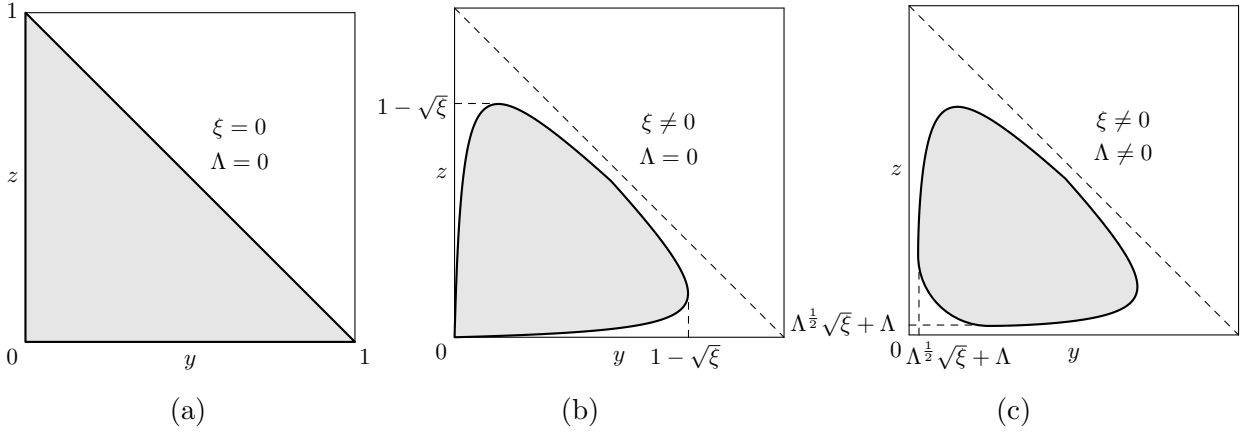
\centerline{\caption Figure 1. Three-particle phase space.}

\bigskip
\noindent
With this parameterization the three-particle phase space simplifies
considerably, namely one finally obtains
$$d\PS\sim dy\,dz,\eqno(4)$$
and the integration limits introduced in [\PhD] are
$$\eqalignno{
&\left\{\matrix{
                 y_+ &=& 1-\sxi \cr
                 y_- &=& \Lambda^{1/2}\sxi+\Lambda \cr
              }\right.&(5\hbox{a})\cr
&z_{\pm}={2y\over4y+\xi}\left[
1-y-\frac{1}{2}\xi+\Lambda+{\Lambda\over y}\pm
{1\over y}\sqrt{(1-y)^2-\xi}\sqrt{(y-\Lambda)^2-\Lambda\xi}\right],
&(5\hbox{b})\cr}
$$
where the masses squared are denoted as $\xi=4m_i^2/q^2$ with $m_i=m_1=m_2$,
and $\Lambda=m_3^2/q^2$. Hence, the cross sections reduce to the
following structure
$$\int\!\!d\PS\,\sum_i f_i\sim\sum_i\int\!\!dy\,dz\,f_i(y,z),\eqno(6)$$
where the individual $f_i(y,z)$ are all simple rational functions in $y$ and $z$.
We have found that if differential and doubly polarized states are
considered~[\NPB], these double integrals generalize to the types $(m,n)$
$$\int\!\!dy\,dz
{f_i(y,z)\over\left[(1-y)^2-\xi\right]^{m/2}
              \left[(1-z)^2-\xi\right]^{n/2}},\eqno(7)$$
where $m,n=0,1,2,3$ independently. The integrals of (7)
will generally display two classes of divergences. The so-called
{\it collinear} singularities manifest themselves as poles and/or logarithmic
singularities in $\xi=0$ and occur on the boundaries of the triangle in
Figure~1a. They can be {\it regularized} by assuming $\xi\neq0$, which deforms
the phase space according to Figure~1b. Here, the so-called {\it soft}
divergences still survive at the origin and take the form of a logarithmic
singularity in $\Lambda$. Assuming $\Lambda\neq0$ finally renders all
intermediate phase-space integrals singularity-free. It is important to
notice that the final cross section (6) is a physical quantity and can
not contain any divergences. Therefore, in the total sum the limits $\xi\to0$
and $\Lambda\to0$ will always yield finite results. The technique of
{\it regularization} is a means of controlling the intermediate
divergences caused by the particular decomposition (7). Typical three-body
reactions in quantum field theory consider two massive particles
and one massless particle, assuming $\xi\neq0$ and $\Lambda=0$.

\bigskip\noindent
{\bf 2. Convergent integrals.}\enspace
For the convergent case, we use as a demonstration of our technique the simplest
integral of the type $(2,1)$:
$$I=\int{dy\,dz\over\left[(1-y)^2-\xi\right]\sqrt{(1-z)^2-\xi}}=
\int\limits_{y_-}^{y_+}{dy\over(1-y)^2-\xi}\left.
\ln\left(\sqrt{(1-z)^2-\xi}+z-1\right)\right|^{z_+}_{z_-},\eqno(8)$$
which contains the phase-space boundaries (5) with $\Lambda=0$.
There is no pole at $(1-y)^2-\xi=0$ as the individual contributions $I_+$ and $I_-$,
corresponding to the limits $z_+$ and $z_-$,
have similar asymptotic behavior with opposite signs and cancel each other.
However, for a separate calculation of these parts, we require
a spurious cut $0<\eps\ll1$:
$$I=I_+ + I_- \quad\hbox{with}\quad
I_\pm=\pm\int\limits_0^{1-\sxi-\eps^2}
{dy\over(1-y)^2-\xi}
\ln\left(\sqrt{(1-z_\pm)^2-\xi}+z_\pm-1\right).\eqno(9)$$
Next, we consider the substitutions
$$k_\pm=1-y\pm\sqrt{(1-y)^2-\xi}\eqno(10\hbox{a})$$
for $I_\pm$, respectively, which gives the following integration limits at
the singular points:
$$k_\pm=\sxi\pm\tilde\eps+\mathcal O(\tilde\eps^2)
\quad\hbox{with}\quad
\tilde\eps=\sqrt{2}\xi^{1\over4}\eps.
\eqno(10\hbox{b})$$
Applying (10) in (9) and using the new variable $v=\sqrt{1-\xi}$ in the limits,
one gets
$$\eqalignno{
I_+=&{1\over\sxi}\int\limits_{\sxi+\tilde\eps}^{2-\sxi}dk
\left[{1\over k-\sxi}-{1\over k+\sxi}\right]\Big\{\ln\xi-\ln(2-k)\Big\} \cr
    &\quad + {1\over\sxi}\int\limits_{2-\sxi}^{1+v}dk\left[
{1\over k-\sxi}-{1\over k+\sxi}\right]\ln(2-k),
&(11\hbox{a})\cr
I_-=&{1\over\sxi}\int\limits_{1-v}^{\sxi-\tilde\eps}dk\left[
{1\over k-\sxi}-{1\over k+\sxi}\right]\Big\{\ln\xi-\ln(2-k)\Big\}.
&(11\hbox{b})\cr
}$$
Integrals (11) can now be solved without further difficulty in terms of
real dilogarithms~[\Lewin]
$$\dilog(x)=-\int\limits_0^x\!\!d\rho\;{\ln(1-\rho)\over\rho}\,.\eqno(12)$$
Especially for more complicated integrals of the general type $(m,n)$, we use
repeated cyclic application of dilogarithmic identities (see also~[\Lewin])
to dramatically simplify the final solutions.
The final result is:
$$\eqalignno{
I =
& 2\dilog\left({\sxi\over2+\sxi}\right)-2\dilog\left({\sxi\over2-\sxi}\right)
  -\dilog\left({1-v\over2+\sxi}\right)+\dilog\left({1-v\over2-\sxi}\right)
  -\dilog\left({2-\sxi\over1+v}\right) \cr
& -\dilog\left({1+v\over2+\sxi}\right)
  +\frac{1}{4}\ln\left({1+v\over1-v}\right)\left[\,-\lxi+2\ln(2-\sxi)
  -\frac{1}{2}\ln\left({1+v\over1-v}\right)\,\right] \cr
& +\frac{1}{2}\lxi\Big[-\frac{1}{4}\lxi+\ln(2-\sxi)\,\Big]
  -\frac{1}{2}\ln^2(2-\sxi)+{\pi^2\over3}. 
& \hfill (13)\cr
}$$
It shall be noted, that rigorous numerical tests have confirmed the validity
of our method. Full analytic results of integrals of this type have been
applied in various contexts (see {\it e.g\/} Refs.~[\PhD,\NPB] and work in
progress).

\bigskip\noindent
{\bf 3. Divergent integrals.}\enspace
For some pathological cases the intermediate phase-space integrals will diverge
because of additional poles in the integrand functions $f_i$ in (7). Then, one
needs to regularize with $\Lambda\neq0$ to avoid contact with the origin as shown
in Figure~1c. A more intricate example is $f=1/z^2$ for integral type~$(2,1)$:
$$J=\int{dy\,dz\over\left[(1-y)^2-\xi\right]\sqrt{(1-z)^2-\xi}}
\,{1\over z^2}\eqno(14)$$
Because of the phase-space symmetry
$y\leftrightarrow z$, it is easy to see that this integral takes the form
$$J=
-{1\over2\sxi}\int\limits_{y_-}^{y_+}{dy\over\sqrt{(1-y)^2-\xi}}
{1\over y^2}\left.
\ln\left(-{z-1+\sxi\over z-1-\sxi}\right)
\right|^{z_+}_{z_-}\eqno(15)$$
with the complicated phase-space boundaries~(5) for $\Lambda\neq0$. For $|z|\ll1$,
we use the expansion
$$\ln\left(-{z-1+\sxi\over z-1-\sxi}\right)=
\ln\left(-{1-\sxi\over1+\sxi}\right)-{2\sxi\over v^2}z+\mathcal O(z^2)\eqno(16)$$
to split the integral into
$$J=
\underbrace{{1\over v^2}\int{dy\,dz\over\sqrt{(1-y)^2-\xi}}
            {1\over y^2}}_{\displaystyle J_1}
+\underbrace{\int\limits_{y_-}^{y_+}{dy\over\sqrt{(1-y)^2-\xi}}{1\over y^2}
\left[-{1\over2\sxi}\ln\left(-{z-1+\sxi\over z-1-\sxi}\right)-{1\over v^2}z
\right]^{z_+}_{z_-}}_{\displaystyle J_2},
\eqno(17)$$
where the second integral $J_2$ is rendered convergent. Therefore, we can
safely put $\Lambda=0$ in this term and proceed as outlined in Section~2
(with {\it two} spurious cuts, one for the upper and one for the lower
integration limit). Straightforward computation yields:
$$\eqalignno{
J_2 =
& \dilog\left(-\sqrt{1+v\over1-v}\right)-\dilog\left(-\sqrt{1-v\over1+v}\right)
  +2\left[\dilog\left(\sqrt{1-v\over1+v}\right)
         +\dilog\left(2{1-\sxi\over1+v-\sxi}\right) \right. \cr
& \left. -\dilog\left(2{1-\sxi\over1-v-\sxi}\right) 
         -\dilog\left({2\over1+v+\sxi}\right)
         -\dilog\left({1-v+\sxi\over2}\right) \right] \cr
& +\frac{1}{8}\left[\lxi-2\ln\left(1-\sxi\right)-12\ln v
    +\frac{1}{2}\ln\left({1+v\over1-v}\right)-2\ln2\right]
    \ln\left({1+v\over1-v}\right) \cr
& +\frac{1}{4}\left[-\frac{1}{4}\lxi-\ln\left(1+\sxi\right)
    +{8v\over\xi}+\ln2\right] \lxi 
  -\frac{1}{2}\left[\frac{1}{2}\ln\left(1+\sxi\right)
    -\ln2\right] \ln\left(1+\sxi\right) \cr
& +{4v\over\xi}\left[-\lxi+2\ln\left(2-\sxi\right)
    +\frac{1}{2}\ln\left({1+\sxi\over1-\sxi}\right)-\ln2+1\right]
    -\frac{1}{4}\ln^22+{\pi^2\over3}\,.
& (18) \cr
}$$
The splitting procedure (17) works for all known cases using adequate expansions
similar to (16).  Its main objective is to isolate the logarithmic divergences
in $\Lambda$ and reduce the difficult divergent part of the integrals to
integrals of lower type. Here, the first integral of (17) is of the lower
type~$(1,0)$:
$$J_1=\int{dy\,dz\over\sqrt{(1-y)^2-\xi}}{1\over y^2}.\eqno(19)$$
By integrating over $z$ and using (5b), it follows that
$$J_1=
\int{dy\over\sqrt{(1-y)^2-\xi}}{1\over y^2}\Big(z_+-z_-\Big) \;=\;
4\int\limits_{\Lambda^{1\over2}\sxi+\Lambda}^{1-\sxi}{dy\over y^2}
{\sqrt{(y-\Lambda)^2-\Lambda\xi}\over4y+\xi},\eqno(20)$$
which again can be solved by adding and subtracting a splitting term
to the integrand. The appropriate splitting term is
$\sqrt{(y-\Lambda)^2-\Lambda\xi}/\xi$. By neglecting finite
pieces in $\Lambda$, one obtains
$$J_1={4\over\xi}
\int\limits_{\Lambda^{1\over2}\sxi+\Lambda}^{1-\sxi}{dy\over y^2}
\sqrt{(y-\Lambda)^2-\Lambda\xi}+
4\int\limits_0^{1-\sxi}{dy\over y}\left[
{1\over4y+\xi}-{1\over\xi}\right]
+\mathcal O(\Lambda).\eqno(21)$$
The integral form (21) can now be solved by standard integration techniques
({\it e.g.\/} letting $t=y-\Lambda$ in the first integral). Further expansion
in $\Lambda$ finally produces
$$J_1=
{4\over\xi}\left[
-\ln\Lambda^{1\over2}+\frac{1}{2}\ln\xi+\ln2+\ln\left(1-\sxi\right)
-2\ln(2-\sxi)-1+\mathcal O(\Lambda)\right]\eqno(22)$$
so that $J=J_1+J_2$ in (14) is fully determined.

\bigskip\noindent
{\bf 4. Conclusion.}\enspace
We have outlined a general integration method to obtain closed-form
analytic solutions for complicated dilogarithmic double integrals that
emerge in applications such as perturbative quantum field theory.
Our method combines suitable substitutions, repeated splitting and
isolation of divergent integral parts, and cyclic usage of dilogarithmic
identities.

In particular, with this approach it was possible to calculate fully 
analytic predictions for reactions in particle physics~[\PhD,\NPB],
where in the past only numerical estimates were available. It also
permitted to find simple and highly accurate polynomial approximations
of the Schwinger type for these processes~[\NPB].

The method has been successfully implemented on computer algebra systems
to help create extensive tables of relevant integrals of type $(m,n)$.
They form the basis of numerical C/FORTRAN libraries that will be
available in the near future.

\bigskip\noindent

\centerline{\bf References}
\medskip
\bib
[\Byckling]
E.~Byckling and K.~Kajantie, {\it Particle Kinematics},
John Wiley \& Sons, New York, (1973).

\bib
[\Sterman]
G.~Sterman, {\it An Introduction to Quantum Field Theory},
Cambridge University Press, Cambridge, 98--106, (1993).

\bib
[\PhD]
M.M.~Tung, {\it Dimensional Reduction Methods in Perturbative QCD,}\/
Ph.D.\ Thesis, University of Mainz, 65--92, (1993);
J.G.~K\"orner, A.~Pilaftsis and M.M.~Tung,
One-loop QCD mass effects in the production of polarized bottom and
top quarks, {\it Z.\ Phys.\ \bf C63}, 575--579, (1994).

\bib
[\NPB]
M.M.~Tung, J.~Bernab\'eu and J.~Pe\~narrocha, $O(\alpha_s)$ spin-spin
correlations for top and bottom quark production in $e^+ e^-$ annihilation,
{\it Phys.\ Lett.\ \bf B418}, 181--191, (1997), and references therein.

\bib
[\Lewin]
L.~Lewin, {\it Dilogarithms and Associated Functions},
MacDonald, London, (1958);
L.~Lewin, {\it Polylogarithms and Associated Functions},
North-Holland Publishing, New York, (1981).

\bye